\documentstyle[prl,aps,psfig,twocolumn,floats]{revtex}
\input psfig
\psfull
\draft

\newcommand{\al}{\alpha}

\newcommand{\De}{\Delta}
\newcommand{\ep}{\epsilon}

\newcommand{\si}{\sigma}

\newcommand{\ra}{\rightarrow}
\newcommand{\be}{\begin{equation}}
\newcommand{\ee}{\end{equation}}
\newcommand{\bea}{\begin{eqnarray}}
\newcommand{\eea}{\end{eqnarray}}
\newcommand{\bean}{\begin{eqnarray*}}
\newcommand{\eean}{\end{eqnarray*}}

\newcommand{\gsim}{\stackrel{>}{\sim}}
\newcommand{\lsim}{\stackrel{<}{\sim}}

\title{ Cosmic Microwave Background Anisotropies
 from Scaling Seeds: \\ Fit to Observational Data} 
\author{R. Durrer$^1$ \and M. Kunz$^1$  \and C. Lineweaver$^2$
	\and M. Sakellariadou$^1$}
\address{$^1$D\'epartement de Physique Th\'eorique,
	 Universit\'e de Gen\`eve,
	24 quai Ernest Ansermet, CH-1211 Gen\`eve 4, Switzerland}
\address{$^2$Observatoire de Strasbourg, 11 rue de l'Universit\'e, 
67000  Strasbourg, France}
\begin{document}
\maketitle
\begin{abstract}
We compute cosmic microwave background angular power spectra
for scaling seed models of structure formation. A generic 
parameterization  of the energy momentum tensor of the seeds
is employed. We concentrate on two regions of parameter space inspired by
 global topological defects: ${\cal O}(4)$ texture models and 
the large-$N$ limit of ${\cal O}(N)$ models.
We use $\chi^{2}$ fitting to compare these models to recent 
flat-band power measurements of the cosmic microwave background.
Only scalar perturbations are considered.
\end{abstract}

Inflation and topological defects are  
two families of models to explain the origin of large scale structure in 
the universe.
In models with topological defects or other types
of seeds, fluctuations are  generated continuously and evolve
according to inhomogeneous linear perturbation equations. 
Seeds are  any   non-uniformly distributed form of energy, which 
contributes only a small fraction to the total energy density of the universe
and which interacts with the cosmic fluid only gravitationally.
We are particularly interested in global topological defects 
playing the role of seeds.  

Cosmic microwave background~(CMB) anisotropies provide a link 
between theoretical predictions and observational data, which may allow us
to distinguish between inflationary perturbations and defects, by purely
linear analysis.   On large angular scales, both families of models 
predict an approximately scale-invariant Harrison-Zel'dovich spectrum
\cite{h,z}. For inflationary models this can be seen analytically. 
Scale-invariance for defects was discovered numerically
\cite{SPT,DZ,Shellard}; simple analytical arguments are given 
in \cite{Roma}.
At small angular scales ($0^{\circ}\!\!.2 \lsim \theta \lsim 1^{\circ}$), 
the predictions of inflation and defects
are different. CMB observations at these scales may soon be 
sensitive enough to distinguish the two families of models.
%\vspace{0.2cm}

In a recent work \cite{MR}, two of us investigated the general behavior
of CMB anisotropies induced by seeds. 
Here, for simplicity, we restrict ourselves to scalar type
perturbations. Thus, the models presented in this work are not close
approximations to the ${\cal O}(4)$ texture model
for which the Sachs-Wolfe~(SW) plateau is dominated by vector and 
tensor modes \cite{DGS,Seljak}.

This restriction does not render our work uninteresting. There
may very well be models with scaling seeds leading to small vector and
tensor contributions, e.g., due to symmetry constraints (spherical
symmetry) or in models with non-relativistic seeds. Here, we
assume a completely phenomenological standpoint: we 
investigate whether models with purely scalar seeds can reproduce the
data. In subsequent work we plan to study how severely vector and
tensor contributions are restricted by the data. 
However, the models already excluded on the basis of our work, will not be 
resurrected once vector and tensor modes are included. 

In our models, we characterize the energy momentum tensor of the source by
four seed functions which we term $f_\rho,~ f_p,~f_v,$ and $f_\pi$, defined
by (see \cite{d90,RuthReview})
\begin{eqnarray}
\Theta_{00}&=&M^2f_{\rho} \\ 
\Theta^{(s)}_{i0}&=&M^2f_{v,i} \\
\Theta^{(s)}_{ij}&=&M^2[\{f_p-(1/3)\Delta f_{\pi}\}\gamma_{ij}+ f_{\pi|ij}] ~,
\label{seed}
\end{eqnarray}
where $\Delta$ denotes the Laplacian and $_|$ is the covariant derivative
with respect to the metric $\gamma$ of  three space. $M$ is a 
typical ``mass'', or energy scale, of the 
seeds. The gravitational strength of the seeds is characterized by the
dimensionless parameter $\ep=4\pi GM^2$.
The superscript $^{(s)}$ indicates that only the scalar contribution to 
$\Theta_{i0}$ and $\Theta_{ij}$ is included here. 
Since seeds interact with other matter
components only gravitationally, the seed functions satisfy the
covariant conservation equations  \cite{d90} 
\begin{eqnarray}
\dot f_{\rho}-\Delta f_v +(\dot a/a)(f_{\rho} + 3 f_p)&=&0\label{cons1}\\
\dot f_v +2(\dot a/a) f_v-f_p-(2/3)\Delta f_{\pi}&=&0~,
\label{cons2}
\end{eqnarray}
where $a$ is the scale factor and dot stands for derivative with
respect to conformal time $t$.

We define ``scaling seeds'' to be seeds for which the power spectra
$\langle|f_{\bullet}|^2\rangle(k,t)$ are, up to an overall power of
$t$ determined by dimensional reasons, functions  of $x=kt$ only. 
Thus, the power spectra of the functions $f_{\bullet}$ are of
the form
\begin{eqnarray}
\langle|f_{\rho}|^2\rangle &=&t^{-1}~F_1^2(x)\nonumber\\
\langle|f_p|^2\rangle &=&t^{-1}~F_2^2(x)\nonumber\\
\langle|f_v|^2\rangle &=&t~F_3^2(x)\nonumber\\
\langle|f_{\pi}|^2\rangle &=&t^3~F_4^2(x)~.
\label{f2}
\end{eqnarray}
Furthermore, we require that the seeds
decay on sub-horizon scales. This behavior
is found in simulations for the seed functions of 
global textures and is also supported by the
large-$N$ limit of global ${\cal O}(N)$ models \cite{KD,KD2}. Simple
analytical arguments indicate that all types of models with scaling 
seed functions which decay fast enough inside the horizon lead to a
Harrison-Zel'dovich spectrum\cite{Arg}. 

Numerical simulations of global ${\cal O}(N)$ models show 
that on super-horizon scales ($x\ll 1$), 
$\Theta_{ij}$ and $\Theta_{00}$ have
white noise spectra , whereas $|\Theta_{i0}|^2$
behaves like $k^2$.  Furthermore,
the power spectra of the functions $f_{\bullet}$ do not 
depend on the direction of {\bf k}. Thus the $F_i$'s are even 
functions of $x=kt$. Consequently, $F_1, F_2, F_3 \ra $~const., 
while $F_4\propto 1/x^2$ for $x\ra 0$.
Since the energy momentum tensor of the seeds decays inside the
horizon, $F_i\rightarrow 0$ for $x\rightarrow \infty$. 
In this work we approximate the random variables $f_{\bullet}$ by
the square root of their power spectra.
Motivated by numerical simulations and the considerations described above 
we model the functions $F_{1}, F_{3}$ as 
\be
F_1 = {A_1\over 1+\al_1x^{n_1}} ~~~,~~~~~
F_3    = {A_3\over 1+\al_3x^{n_3}}~.
\ee
$F_2$ and $F_4$ are then given by energy momentum
conservation, Eqs.~(\ref{cons1}, \ref{cons2}).

The gravitational action of the seeds is determined by the induced Bardeen
potentials, which are not only due to the seeds but also contain
contributions from the matter and radiation fluids. 
Once the fluid perturbations and the Bardeen potentials are
determined, one calculates CMB anisotropies by standard
methods for each model parametrized by
 $(A_1,\al_1,n_1,A_3,\al_3,n_3)$. For details see Ref.~\cite{MR}.

In this letter we present the results of a parameter study and we fit to
the observational data available. We compare the anisotropy power
spectra  obtained in our models with observations.
The cosmological parameters used are
$h=0.5, \Omega_B=0.0125 h^{-2}, \Omega=1$ and  $\Lambda=0$.
We are thus considering scaling seed models in the context of flat 
cold dark matter universes. 

We investigate two types of models. 
In the first one we choose $n_1=2,~n_3 =4$; a choice supported by numerical 
simulations of global textures \cite{DZ}; we refer 
to this first type of scaling seed models as  model A.
In the second one we set $n_1=5/2$ and $n_3=7/2$; a result 
obtained analytically in the large-$N$ limit of ${\cal O}(N)$ 
models\cite{KD}. We call it  model B. 

We set the arbitrary normalization by fixing $A_{1}=1$ and we vary
$A_{3} (= A_{3}/A_{1})$. To make the 
calculations feasible, we further reduce the remaining 3-dimensional
parameter space $(A_{3}, \alpha_{1}, \alpha_{3})$ to 2-dimensions
$(A_{3}, \alpha_{1})$, by setting $\alpha_{3} = \alpha_{1}/2$.
Thus, our fits are displayed as contour plots in the
$\alpha_{1} - A_{3}$ plane (see Figs.~3 and 4). We have also 
investigated $\alpha_{3} = 2\alpha_{1}$ and obtained qualitatively
similar results.

The $A_3$ dependence is similar for both types of
models. 
For $A_3 \lsim 0.06$, the relative amplitude of the acoustic peak, with respect
to the SW plateau, decreases as $A_{3}$ decreases (see Figs.~1a and c).

There is a
particular value of the constant $A_3$ for which the coefficient of
the $1/x^2$ term in $F_4$ vanishes. For $x\ll 1$, one obtains from 
Eqs.~(\ref{cons1},\ref{cons2}) 
$F_4\approx A_4/x^2$ with $A_4=-(3/8)A_1(1+18A_3/A_1)$
in the matter dominated era. If 
$A_3 = (-1/18) A_1\equiv A_{crit}\sim -0.06$, we thus find
$A_4= 0$ \cite{MR}. This $1/x^2$ term dominates on super-horizon
scales and (for big enough $\al_i$) determines the amplitude of
the SW plateau. Its absence is thus expected to lead to a  
higher relative amplitude of the first acoustic peak, which is well
visible in Figs.~1a and 1c.

\begin{figure}[htb]
\vspace{0.3cm}
\centerline{\psfig{figure=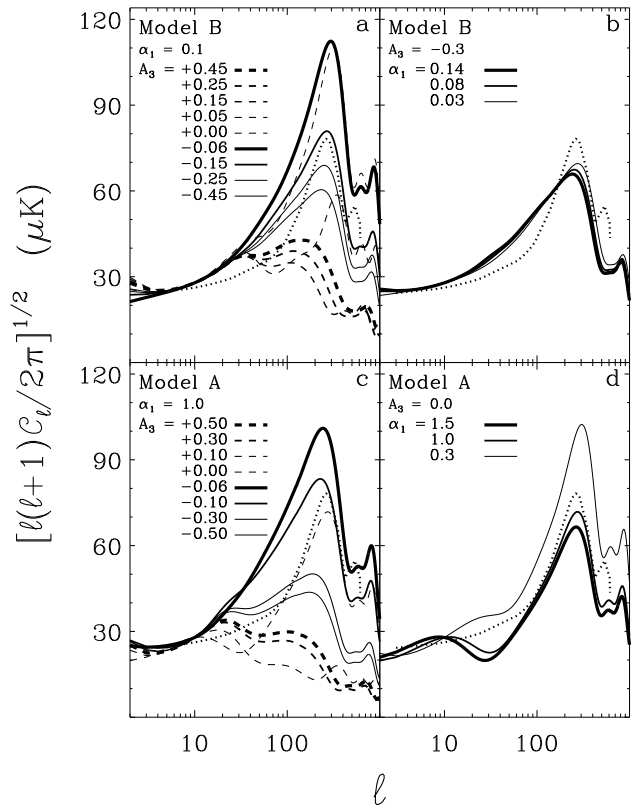,width=8.3cm}}
\caption{Parameter dependence of the calculated
power spectra. To illustrate the dependence on $A_3$, we choose values
around $A_{crit} = -0.06$ (see text). The dotted line is a polynomial
fit to the data and is the same in each panel as a reference.}
\end{figure}

\begin{figure}[htb]
\centerline{\psfig{figure=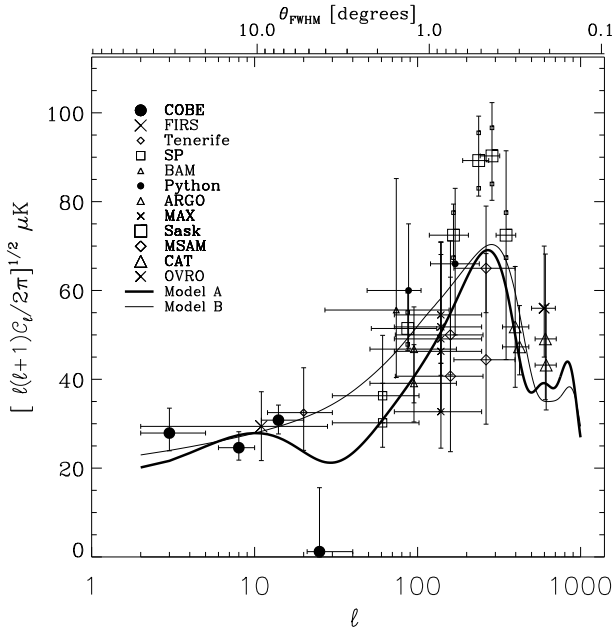,width=8.3cm}}
\label{fig:data}
\caption{Flat-band power measurements used in this analysis.
The best-fit models of two scaling seed models are also shown:
one most relevant to  model A (Fig.~3) and the other 
for  model B (Fig.~4).}
\end{figure}

In our exploration of parameter space we vary 
$-1.0 \le A_{3} \le +1.0$ for  model A and
$-1.2 \le A_{3} \le +0.5$ for model B.
For $\alpha_{1}$ we choose the parameter range 
$0.01 \le \alpha_{1} \le 2.0$ for  model A and 
$0.001\le \alpha_{1}\le 0.141$
for  model B.
We normalize the power spectra at $\ell=10$ by fitting to the data. 

We use $\chi^{2}$ fitting to compare our models to recent
flat-band power measurements of the CMB.
The method and a compilation of the data is described in detail in 
\cite{Line97}.
The most recent data and improvements to the $\chi^{2}$ method are described
in \cite{LB97a,LB97b}.

Fig.~2 plots the data used along with the best-fit power
spectra for  the two types of models described above.
These best-fit models are indicated by the ``X'' in Figs.~3 and 4.
Fig.~3 is a contour plot in the $\alpha_{1}- A_{3}$ plane for
model A while Fig.~4 is the analogous one for model B.

There are 32 data points and 28  degrees
of freedom (28 = 32 - 2 (fitted defect parameters) - 1 (normalization) 
- 1 (sliding Saskatoon absolute calibration)).
Model A yields a $\chi^2$ minimum value 
$ \chi_{\min}^2 = 21.2$,  while  model B yields
$ \chi_{\min}^2 = 27.1$. Thus the fits are good.  
\begin{figure}[htb]
\centerline{\psfig{figure=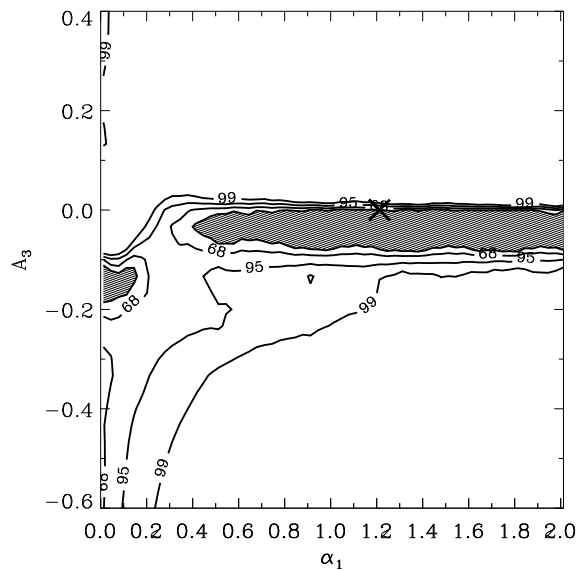,width=7.5cm}}
\label{fig.fit1}
\caption{$\chi^{2}$ fit of  model A to the CMB data.
The ``${\bf X}$'' marks the $\chi^{2}_{min}$; the grey area is
the $(\chi^{2}_{min} + 1)$, $\sim 1\sigma$-likelihood region.
Contours marked ``68'', ``95'' and ``99'' refer to goodness-of-fit contours.
For example, under the assumption that the errors on the flat-band power
measurements are Gaussian, the probability of obtaining a 
$\chi^{2}$ value less than the
value obtained on the ``95'' contour is $95\%$.}
\end{figure}

To interpret our $\chi^2$ fits, we first note the following: if
perturbations decay fast enough, the height of the first acoustic peak
is determined by $f_\rho+3f_p \sim A_1/4$, while the SW plateau is
fixed by $A_4$. The $\chi^2$ is then not expected to be very sensitive
to $\alpha$. This indeed is seen in  model B for 
$\alpha_1\gsim 0.03$ (Fig.~4 and also Fig.~1 b) and in 
 model A for $\alpha_1\gsim 0.4$ (Fig.~3.).

We also have investigated the dark matter power spectra for some of
the parameter space of our models.
For the best fitting models we obtain $\si_8 \sim 0.6\pm 0.2$ which
is in reasonable agreement with observations. Values of $(A_3,\al_1)$
which are excluded from the $C_{\ell}$ analysis, often lead to far too
small values of $\si_8$. We found, however, that the bend in the power spectrum
of our ``best models'' lies at somewhat smaller scales than in the analysis
of APM and IRAS galaxies done by Peacock \cite{peacock}.  
Since his analysis assumes Gaussian statistics, we are very reluctant
to draw any strong conclusions from this comparison.
 
In the matter dominated era, numerical simulations of ${\cal O}(4)$-textures
give  $A_1\sim 4,\alpha_1\sim 0.012$ and 
$A_3\sim 0.37, \alpha_3\sim 0.05$ \cite{KD2}. 
For this model, the scalar contribution
to the SW plateau is $\sim 1.1 \ep^2$ at $\ell\approx 10$ 
and the height of the first
acoustic peak is about $\sim 5 \ep^2$. On the other hand, full simulations,
which include vector and tensor modes\cite{DZ}, lead to an
amplitude of the SW contribution on the order of
$\ell(\ell+1)C_{\ell}\sim 8^{+4}_{-2}\ep^2$. Therefore the vector and  
tensor parts contribute more than $\sim 80\%$ to the SW  plateau, while
they are not expected to influence the acoustic peaks. 
In the full texture models, the acoustic peaks are thus expected to be
substantially too low to fit the data. This result was pioneered 
in \cite{DGS} and has
now been confirmed by full numerical simulations \cite{Seljak}. 
In Ref.~\cite{Seljak} decoherence \cite{decoh}
 has also been taken into account, which
further reduces the acoustic peaks without influencing substantially
the SW plateau. 

The fact that we obtain a parameter range compatible
with currently available observational data is clearly not in 
contradiction with the result of \cite{DGS} and \cite{Seljak}, namely
that the acoustic peaks for the conventional ${\cal O}(4)$-texture 
model are very small or even completely absent. 

\begin{figure}[htb]
\centerline{\psfig{figure=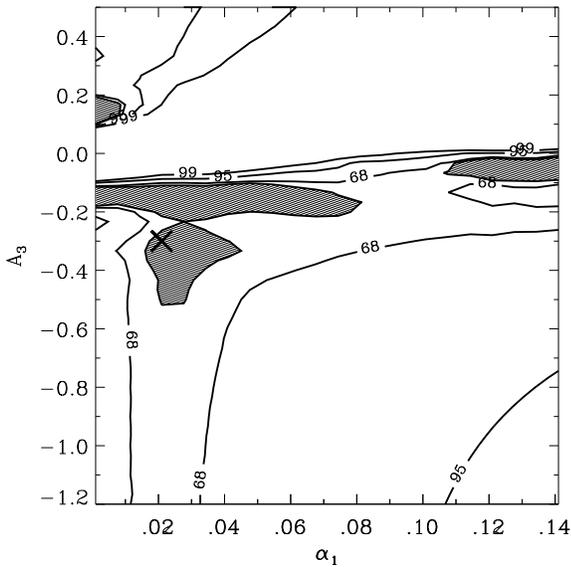,width=7.5cm}}
\label{fig:largeNcontours}
\caption{Same as previous figure except here we are fitting 
 model B to the data.}
\end{figure}

Fig.~3 shows that small negative values of $A_3$ ($-0.1\le A_3\le 0$)
are preferred. For $\al_1\gsim 0.4$ the result does not depend strongly
on $\al_1$. For smaller values of $\al_1$, ($\al_1\lsim 0.2$) somewhat
larger values of $|A_3|$ are preferred since the acoustic peaks 
are already enhanced due to the slower decay of the seed functions.

Fig.~4 is the  model B analog of Fig.~3. Also here,
a positive $A_3$ is excluded for $\al_1\gsim 0.01$. A value of
$A_3\sim -0.1$ is generically preferred. Note however, that the $\chi^2$
``landscape'' within the parameter range explored in Figs.~3,4 is rather flat,
and values within the ``68'' contour are reasonably compatible with current
data.\vspace{0.2cm}

In this letter, our aim was not to test whether a given model with
topological defects can fit the data. We wanted to
investigate, whether present observations of CMB anisotropies can 
already rule out a generic class of seed perturbations constrained just by
energy momentum conservation and scaling arguments. 
Our analysis indicates that the answer to this question is {\sl no}.
 
As a continuation of this work, we plan to  
include vector and tensor perturbations as well as decoherence 
in our models. 
We also want to study whether there are more severe restrictions on 
defect models than just energy momentum conservation and scaling;
for example, to see whether the vector component always dominates the 
level of the SW plateau.
\vspace{0.2cm}

{\bf Acknowledgment}\hspace{0.5cm}
This work is partially supported by the Swiss NSF.  
M.S. acknowledges financial support from the Tomalla foundation.
C.H.L acknowledges NSF/NATO post doctoral fellowship 9552722.


\begin{thebibliography}{99}
\bibitem{Arg} For models with decaying seeds, the Bardeen potentials
are roughly constant inside the horizon. Therefore, the SW contribution is
given by $(\De T/T)_{\ell}|_{SW} \sim (\Phi -\Psi)(k,x_{dec})j_{\ell}(x_0) 
+\int_{x_{dec}}^1({\Phi}'-{\Psi}')(k,x)j_{\ell}(x_0-x)dx$, where
$x=kt$ and prime stands for derivative w.r.t. $x$. 
The lower boundary of the integrated term roughly cancels the
ordinary SW contribution and the upper boundary leads, by simple dimensional
reasons, to a Harrison-Zel'dovich spectrum.
\bibitem{decoh}
	Decoherence is  the decay of the correlation functions, 
        $\langle f_i(t)f_j(t')\rangle$, 
        approximated  here simply  by $\langle f_i(t)f_j(t')\rangle 
        \sim \sqrt{\langle |f_i(t)|^2\rangle\langle|f_j(t')|^2\rangle}$. This
        is likely to influence the structure of the
        acoustic peaks.
\end{thebibliography}
\end{document}